\documentstyle[aps,epsfig]{revtex}
\newcommand{\ds}{\displaystyle}
\newcommand{\bl}{\bigl}
\newcommand{\br}{\bigr}
\newcommand{\beqn}{\begin{eqnarray}}
\newcommand{\eeqn}{\end{eqnarray}}
\newcommand{\beq}{\begin{equation}}
\newcommand{\eeq}{\end{equation}}
\newcommand{\barr}{\begin{array}}
\newcommand{\earr}{\end{array}}

\newcommand{\nv}{\not\! v}
\newcommand{\nvpr}{{\not\!v}^{\,\prime}}
\def\qed{\hbox{${\vcenter{\vbox{                        %HOLLOW SQUARE
   \hrule height 0.4pt\hbox{\vrule width 0.4pt height 6pt
   \kern5pt\vrule width 0.4pt}\hrule height 0.4pt}}}$}}
\begin{document}
\twocolumn
%
%\preprint{Preprint Number: \parbox[t]{45mm}{MZ-TH-99-}}
%
\title{On the choice of heavy baryon currents  
in the relativistic three-quark model}
\author{M. A. Ivanov$^{\,1}$, 
        J. G. K\"orner$^{\,2}$,  
        V. E. Lyubovitskij$^{\,1,3,6}$,
        M. A. Pisarev$^{\,3,4}$, 
        A. G. Rusetsky$^{\,1,5,7}$ 
                                             \vspace*{0.2\baselineskip}}
\address{
$^1$Bogoliubov Laboratory of Theoretical Physics, 
Joint Institute for Nuclear Research, 
141980 Dubna (Moscow region), Russia \vspace*{0.2\baselineskip} \\
$^2$Johannes Gutenberg-Universit\"at, Institut f\"ur Physik,
Staudinger Weg 7, D-55099 Mainz, Germany \vspace*{0.2\baselineskip} \\
$^3$Department of Physics, Tomsk State University, 
634050 Tomsk, Russia \vspace*{0.2\baselineskip} \\ 
$^4$University Center of the Joint Institute for Nuclear Research, 
141980 Dubna (Moscow region), Russia  \vspace*{0.2\baselineskip} \\ 
$^5$HEPI, Tbilisi State University, 380086 Tbilisi, Georgia
\vspace*{0.2\baselineskip} \\
$^6$Institut f\"ur Theoretische Physik, Universit\"at T\"ubingen, 
Auf der Morgenstelle 14, D-72076 T\"{u}bingen, Germany
\vspace*{0.2\baselineskip} \\
$^7$Institute for Theoretical Physics, University of Bern,
Sidlerstrasse 5, CH-3012, Bern, Switzerland,\vspace*{0.2\baselineskip}}
\date{Preprint MZ-TH/99-51; 
}
\maketitle
\begin{abstract}
We test the sensitivity of bottom baryon observables with regard to 
the choice of the interpolating three-quark currents  within the
relativistic three-quark model. We have found that the semileptonic
decay rates are clearly affected by the choice of currents, whereas the
asymmetry parameters show only a very weak dependence on the choice of
current.
\end{abstract}
\vspace*{\baselineskip}
%
%\pacs{PACS number(s): 11.10.St, 12.39.Ki, 13.30.Ce, 14.20.Lq, 14.20.Mr}
%
\section{Introduction}
The forthcoming experimental data on exclusive bottom baryon decays call for  
a comprehensive theoretical analysis of their spectra and their decay 
properties. During the last decade heavy baryon 
transitions have been studied in detail within the Heavy Quark Effective 
Theory employing QCD sum rule methods or nonrelativistic and relativistic 
quark models, etc. (see, for example, the reviews in \cite{KKP,MN} and 
the papers \cite{IW,HKKG,MRR,EIKL,H,KM,Yan,Grozin,Cheng,HDH,Groote,ILKK,IKLR1,IKLR2,IKLR3,IKLR4,TOK,DFNR,Dai,SQM,IMF1,IMF2,Sadzik_Zalew,Gupta,Lee,Skyrme}). 
The mass spectrum of heavy baryons as well as their exclusive and inclusive 
decays have been described successfully in these approaches incorporating the 
ideas of QCD. Preliminary results for the $\Lambda_b\to\Lambda_c$ baryon 
Isgur-Wise function and its slope have recently been obtained in Lattice 
QCD \cite{Bowler}.  

In the papers \cite{ILKK,IKLR1,IKLR2,IKLR4,EIL,EI,AIKL,ILL} we proposed and
developed a QCD motivated relativistic three-quark model (RTQM),
which can be viewed as an effective quantum field approach based on an 
interaction Lagrangian of light and heavy baryons interacting with their 
constituent quarks. The coupling strengths of the baryons interacting with 
the three constituent quarks are determined by the compositeness condition 
$Z_H=0$  \cite{EI,SWH} where $Z_H$ is the wave function renormalization 
constant of the hadron. The compositeness condition enables one to 
unambiguously and consistently
relate the theories with both quark and hadron degrees of freedom 
to the effective Lagrangian approaches formulated in terms of hadron
variables only (as, for example, Chiral Perturbation Theory \cite{WGL}
and its covariant extension to the baryon sector \cite{BL}).
Our strategy is as follows. We start with an effective interaction
Lagrangians written down in terms of quark and hadron variables. 
Then, by using  Feynman rules, the $S$-matrix elements describing 
hadron-hadron interactions are given in terms of a set of quark Feynman 
diagrams. 
The compositeness condition serves to avoid double counting of quark 
and hadron degrees of freedom. The RTQM  contains only a few model
parameters: the masses of the light and heavy quarks, and certain scale 
parameters that are related to the size of the distribution of the constituent 
quarks inside the hadron. The RTQM  has been previously used to 
compute the exclusive semileptonic, nonleptonic, strong and electromagnetic
decays of charm and bottom baryons  in the heavy
quark limit $m_Q\to\infty$ always employing the same set of model
parameters \cite{ILKK,IKLR1,IKLR2}.

The objective of this paper is to continue the analysis of heavy baryon 
transitions within the RTQM
\cite{ILKK,IKLR1,IKLR2,IKLR3,IKLR4,EIL,EI,AIKL,ILL}. In particular we shall 
investigate the dependence of heavy baryon observables calculated in the 
RTQM on the choice of three-quark baryon currents. 
In the heavy quark limit there remains a twofold ambiguity in the choice of 
interpolating currents for the ground state baryons. The properties of heavy 
baryons calculated in any model will in general depend on the choice taken for
the baryon currents. It is therefore worthwhile to use a particular model and 
provide a detailed investigation of how the choice of interpolating currents 
affects the outcome of a dynamical calculation. 
For definiteness we shall limit our investigation to $b\to c$ semileptonic 
transitions of $\Lambda$-type and $\Omega$-type heavy baryons 
(such as  $\Lambda_b\to\Lambda_c e \nu_e$, 
$\Sigma_b\to\Sigma_c e \nu_e$, etc.).  

We proceed as follows. First we briefly explain the basic ideas of the RTQM. 
Next we obtain analytic expressions for the heavy baryon Isgur-Wise 
functions and calculate rates and differential distributions in baryonic 
$b\to c$ semileptonic transitions of ground state $\Lambda$-type and 
$\Omega$-type heavy baryons. We compare our numerical results with the 
results of other theoretical approaches. 

\section{Relativistic Three-Quark Model}
We start with a brief description of our approach called 
the Relativistic Three-Quark Model (RTQM). A detailed description of
the RTQM can be found in Refs. \cite{ILKK,IKLR2,AIKL,ILL}.
In the RTQM  baryons are described as bound states of
consitituent quarks. We denote the heavy baryons by  
$B_Q=Qq_1q_2$ which specifies  a bound state of an infinitely large heavy 
quark $Q=b$ or $c$ and two light quarks $q_1$ and $q_2$ with masses $m_{q_1}$ 
and $m_{q_2}$. We express the spatial four-coordinates $(y_i)$ of the 
constituent quarks in terms of the center-of-mass coordinate $(x)$ and the 
relative Jacobi coordinates (see ref. \cite{ILKK}): 
\begin{eqnarray*}  
y_Q&=&x\,,\\
y_{q_1}&=&x+3\xi_1-\sqrt{3}\xi_2\,, \\
y_{q_2}&=&x+3\xi_1+\sqrt{3}\xi_2\,.
\nonumber
\end{eqnarray*}  

The Lagrangian describing the interaction of a heavy baryon $B_Q$
with a single heavy quark $Q$ and two light quarks $q_1$ and $q_2$
simplifies in the heavy quark limit. The Lagrangian can be written 
as \cite{ILKK}
\begin{eqnarray}\label{HB}
{\cal L}_{B_Q}(x)=g_{B_Q}\bar B_Q(x)
J_{B_Q}(x)+{\rm h.c.}
\end{eqnarray}
where $J_{B_Q}$ is a three-quark current with the quantum numbers of the 
heavy baryon given by  

\begin{eqnarray}\label{HB_cur}
J_{B_Q}(x)&=&\Gamma_1
Q^{a_1}(x)\!\int\! d^4\xi_1\!\int\! d^4\xi_2\,
F_{B_Q}(\xi^2_1+\xi^2_2)
\\
&\times& q^{a_2}_1\!\left(x+3\xi_1-\sqrt{3}\xi_2\right)
C\Gamma_2\lambda_{B_Q}
\nonumber\\
&\times&
q^{a_{3}}_{2}\!
\left(x+3\xi_{1}+\sqrt{3}\xi_{2}\right)\varepsilon^{a_1a_2a_3}\,,
\nonumber
\end{eqnarray}

\begin{eqnarray}
&&
F_{B_Q}(\xi^2_1+\xi^2)=
\\
&&
\int\!\frac{d^4k_1}{(2\pi)^4}\!\int\!\frac{d^4k_2}{(2\pi)^4}
e^{-ik_1\xi_1-ik_2\xi_2}{\widetilde{F}}_{B_Q}
\!\left(\frac{k^2_1+k^2_2}{\Lambda^2_{B_Q}}\right).
\nonumber
\end{eqnarray}

Here $\Gamma_1$ and $\Gamma_2$ are appropriate strings of Dirac matrices, 
$\lambda_{B_Q}$ is a flavor matrix and $C=\gamma^0\gamma^2$ is the charge 
conjugation matrix. $F_{B_{Q}}$ is the heavy baryon vertex form factor
defining the momentum distribution of light quarks within the heavy baryon.
 
Unlike the heavy meson case the heavy baryon Lagrangian (\ref{HB}) contains a 
twofold ambiguity in the choice of the spin-flavour structure of the heavy 
baryon currents $J_{B_Q}$ (even in the absence of derivative couplings).
For the $\Lambda$-type baryons ($\Lambda_Q$, $\Xi_Q$) with a light spin zero 
diquark system one has {\it a pseudoscalar current} $J^P_{B_Q}$ and 
{\it an axial current} $J^A_{B_Q}$, both of which have the correct quantum
numbers to serve as interpolating fields for the $\Lambda$-type baryons.
Similarly for the $\Omega$-type baryons
$\Omega_Q$, $\Sigma_Q$ and $\Omega^{\,*}_Q$, $\Sigma^{\,*}_Q$ with 
a spin one diquark system one has a {\it a vector current} $J^V_{B_Q}$ and 
{\it a tensor current} $J^T_{B_Q}$. In terms
of the spinor and flavour structure the two respective currents in each case 
are given by  \cite{Grozin,Groote,ILKK,Shuryak} 
\begin{eqnarray}\label{current}
J^P_{\Lambda_Q}&=&\varepsilon^{abc}Q^au^bC\gamma^5d^c
\\
J^A_{\Lambda_Q}&=&\varepsilon^{abc}\gamma_\mu
Q^a u^bC\gamma^5\gamma^\mu d^c
\\
&&\nonumber\\
J^V_{\Omega_Q}&=&\varepsilon^abc\gamma_\mu\gamma^5
Q^a s^bC\gamma^\mu s^c
\\
J^{T}_{\Omega_Q}&=&\varepsilon^{abc}\sigma_{\mu\nu}\gamma^5
Q^as^bC\sigma^{\mu\nu}s^c
\\
&&\nonumber\\
J^{V;\mu}_{\Omega^{*}_Q}&=&\varepsilon^{abc}
Q^as^bC\gamma^\mu s^c
\\
J^{T;\mu}_{\Omega^{*}_Q}&=&-i\varepsilon^{abc}
\gamma_\nu Q^a s^b C\sigma^{\mu\nu}s^c
\end{eqnarray}
In this paper we investigate the sensitivity of observables on the choice
of heavy baryon currents. We will consider general linear
combinations of the two possible currents for the ground states of heavy 
baryons. Thus we write 
\begin{eqnarray}\label{cur_mix}
J_{\Lambda_Q}\bl(x\br)&=&\alpha_PJ^{P}_{\Lambda_Q}\bl(x\br)
+\alpha_A J^{A}_{\Lambda_Q}\bl(x\br) 
\\
J_{\Omega_Q}\bl(x\br)&=&\beta_VJ^{V}_{\Omega_Q}\bl(x\br)+
\beta_TJ^{T}_{\Omega_Q}\bl(x\br)
\\
J^{\mu}_{\Omega^{*}_Q}\bl(x\br)&=&
\beta_V J^{V;\mu}_{\Omega^{*}_Q}\bl(x\br)+ 
\beta_T J^{T;\mu}_{\Omega^{*}_Q}\bl(x\br)
\end{eqnarray}
Since the $\Omega_Q$ and the $\Omega_Q^*$ are members of the same heavy quark 
symmetry doublet the coefficients $\beta_V$ and $\beta_T$ are the same for 
both. 
As a result of the twofold ambiguity we have to introduce the two additional 
parameters $R_\Lambda=\alpha_A/\alpha_P$ and $R_\Omega=\beta_T/\beta_V$. 

Next we specify our model parameters. 
The heavy-baryon quark coupling constants $g_{B_{Q}}$ are determined by 
the compositeness condition \cite{EIKL,IKLR2,EI,SWH}. The compositeness
condition implies that the renormalization constant of the hadron wave
function is set equal to zero: $Z_{B_Q} = 1 - g_{B_Q}^2
\Sigma^\prime_{B_Q}(M_{B_Q}) = 0$ where
$\Sigma^\prime_{B_Q}$ is the derivative of the baryon mass operator and
$M_{B_Q}$ is the heavy baryon mass.
In Eq. (\ref{HB_cur}) we have introduced a baryon-three-quark vertex
form factor written as $\tilde F_{B_Q}((k_1^2+k_2^2)/\Lambda_{B_Q}^2)$ 
where $\Lambda_{B_Q}$ is a scale parameter defining the distribution 
of the light quarks in the heavy baryon. Any choice of vertex function 
$F_{B_Q}$ is appropriate as long as it falls off sufficiently fast in the 
ultraviolet region to render the Feynman diagrams ultraviolet finite.
In principle, its functional form would be calculable from the solutions
of the Bethe-Salpeter equations for the baryon bound states \cite{IKLR3}
which is, however, an untractable problem at present.
In our previous analysis \cite{AIKL} we found that, using various
forms for the vertex function, the hadron observables
are insensitive to the details of
the functional form of the hadron-quark vertex form factor.
We will use this observation as a guiding principle and choose a simple
Gaussian forms for the vertex function $F_{B_Q}$. 
Its Fourier transform reads \cite{ILKK,IKLR1,IKLR2,ILL}
\begin{equation}\label{BS} 
\tilde F_{B_Q}\left(\frac{k^{2}_{1}+k^{2}_{2}}{\Lambda^{2}_{B_{Q}}}\right)=
\exp\left(\frac{k_1^2+k_2^2}{\Lambda_{B_Q}^2}\right)
\end{equation}
where $\Lambda_{B_Q}$ is a scale parameter defining the distribution of 
$u$ and $d$ quarks in the heavy baryon. For the light quark propagator 
with a constituent mass $m_q$ 
we shall use the standard form of the free fermion propagator
\begin{equation}\label{light}
S_q(k)=\frac{1}{m_q-\not\! k}
\end{equation}
where $m_q=m$ for the $u$ or $d$ quarks (we work in the isospin symmetry 
limit) and  $m_q=m_s$ for the strange quark. 
For the heavy quark propagator we shall use the leading term 
$S_v(k,\bar\Lambda_{q_1q_2})$ in the inverse mass expansion of the free 
fermion propagator: 
\begin{eqnarray}\label{heavy_propagator}
S_Q(p+k)&=&\frac{1} {m_Q-(\not\! p + \not\! k)}\,,
\nonumber\\
&=&
S_v(k,\bar\Lambda_{q_1q_2})+O\left(\frac{1}{m_Q}\right)
\nonumber\\
S_v(k,\bar\Lambda_{q_1q_2})&=&-\frac {(1+\not\! v)}
{2(v\cdot k + \bar\Lambda_{q_1q_2})}
\end{eqnarray}
We  introduce the mass difference parameter 
$\ds{\bar\Lambda_{q_1q_2} = M_{Qq_1q_2} - m_Q}$ which is the difference 
between the heavy baryon mass $M_{Qq_1q_2}\equiv M_{B_Q}$ and the 
heavy quark mass $m_Q$. The four-velocity of the heavy quark is denoted 
by $v$ as usual. As in the light quark propagator we shall neglect a possible 
mass difference between the constituent $u$- and $d$-quark. Thus there are 
altogether three independent mass parameters: $\bar\Lambda$ for heavy 
baryons without strange quarks, 
$\bar\Lambda_s$ for heavy baryons with a single strange quark and 
$\bar\Lambda_{ss}$ for doubly strange  heavy baryons. 

Our set of model parameters are the following: the masses of the light quarks 
$m$ and $m_s$, the vertex scale parameter $\Lambda_{B_Q}$, parameters related 
to the heavy quark propagator  $\bar\Lambda$, $\bar\Lambda_s$ and 
$\bar\Lambda_{ss}$ and the two parameters $R_\Lambda=\alpha_A/\alpha_P$ and 
$R_\Omega=\beta_T/\beta_V$ related to the twofold ambiguity in the choice 
of the heavy baryon currents for the $\Lambda$-type and $\Omega$-type 
baryons. The parameter  $m=420$~MeV has been fixed in Ref~{\cite{ILL}}
from an analysis of nucleon data. 
The parameters $\Lambda_{B_Q}$, $m_s$ and $\bar\Lambda$ are taken from an 
analysis of the $\Lambda_c^+\to\Lambda^0+e^+ + \nu_e$ decay data. A good 
description  of the present average value of the branching ratio 
$B(\Lambda_c^+\to\Lambda e^+ \nu_e)=2.2\%$ can be achieved with 
$\Lambda_{B_Q}=1.8$ GeV, $m_s=570$ MeV and 
$\bar\Lambda=600$ MeV \cite{IKLR4}. In addition, the value of the strange 
quark mass $m_s=570$ MeV gives the best description of the magnetic moments of 
light hyperons ($\Lambda$, $\Sigma$, $\Xi$). The values of the parameters 
$\bar\Lambda_s$ and $\bar\Lambda_{ss}$ are determined from the heuristic 
relations $\bar\Lambda_s=\bar\Lambda+(m_s-m)$ and 
$\bar\Lambda_{ss}=\bar\Lambda+2(m_s-m)$, which gives $\bar\Lambda_s=750$ MeV 
and $\bar\Lambda_{ss}=900$ MeV. Finally, the mass values of the charm and 
bottom baryon states are taken from Ref. \cite{IKLR3} (masses of $\Lambda_Q$, 
$\Xi_c$, $\Sigma_c$ and $\Sigma^*_c$ baryons) and Ref. \cite{ILKK}  
(masses of $\Xi_b$, $\Sigma_b$, $\Sigma^*_b$ and $\Omega_Q$ baryons).  

\section{Results}

\subsection{Matrix elements of semileptonic transitions of heavy baryons} 

\begin{center}
\begin{figure}[t]
\epsfig{file=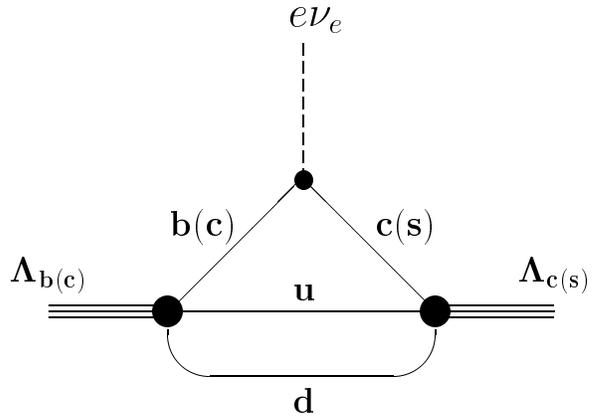,height=6cm}
\caption
{\label{fig1}
Feynman diagram describing the semileptonic 
$b\to c$ decay of heavy baryons $B_b\to B_c + e + \nu_e$ 
}
\end{figure}
\end{center}

The semileptonic $b\to c$ transitions of heavy baryons are described 
by the triangle two-loop quark diagram Fig.~\ref{fig1}.

It takes the following form in the heavy quark limit 
\begin{eqnarray}\label{BC-trans}
&&
\bar u(v')M^\mu(v,v')u(v)=
\\
&&
\frac{G_F}{\sqrt{2}} V_{cb}g^i_{B_Q}g^{f}_{B_Q}
\bar u(v')\Gamma^f_1\!(1+\not\!v')O^\mu (1+\not\! v)
\Gamma^i_1 u(v)\cdot I^{if}_{q_1q_2}
\nonumber\\
&&\nonumber\\
&&
I^{if}_{q_1q_2}(v,v')=
\int\!\frac{d^4k_1}{2\pi^2i}\!\int\!\frac{d^4k_2}{2\pi^2i}
{\widetilde{F}}^{2}_{B_Q}\!\left(\frac{9k^2_1+3k^2_2}{\Lambda_{B_Q}}\right)
\nonumber\\
&\times&
\frac
{
{\rm tr}[\Gamma^f_2S_{q_2}(k_2)\Gamma^i_2S_{q_1}(k_1+k_2)]
}
{ 
\left[k_1v+\bar\Lambda_{q_1q_2}\right]
\left[k_2v'+\bar\Lambda_{q_1q_2}\right]
}
\nonumber
\end{eqnarray}
where $G_F$ is the Fermi weak effective coupling, $V_{cb}=0.04$ is the CKM 
matrix element, $g_{B_Q}^i$ and $g_{B_Q}^f$ are the couplings constants 
of quarks with the initial (i) and the final (f) baryon, respectively. 

The calculational techniques of how to deal with the integral (\ref{BC-trans}) 
can be found in Refs. \cite{IKLR2,IKLR4}. All dimensional parameters in the 
Feynman loop integrals are expressed in units of 
$\Lambda_{B_Q}$. The Feynman integrals are calculated in the Euclidean region 
both for internal and external momenta. The final results are obtained by 
analytic continuation of the external momenta to the physical region after the 
internal momenta have been integrated out. 

After a few steps of  calculation of the  overlap integral 
$\ds{I^{if}_{q_{1}q_{2}}}$ can be written as

\begin{eqnarray}\label{overlap}
I^{if}_{q_1q_2}(v,v')&=&
\int\limits^{\infty}_{0}\!d^{\,4}\alpha
\frac{{\widetilde{F}}^{2}_{B_{Q}}(6z)}{4{\left|A\right|}^{2}}
\\
&\times&
\left\{
m_{q_{1}}m_{q_{2}}{\rm tr}\Bigl[\Gamma^{i}_{2}\Gamma^{f}_{2}\Bigr]
\right.
\nonumber\\
&-&
\left.
m_{q_1}\frac{A^{-1}_{11}+A^{-1}_{12}}{2}
{\rm tr}\Bigl[\Gamma^i_2\bigl(\nv\alpha_3+\nvpr\alpha_4\bigr)
\Gamma^f_2\Bigr]\right.
\nonumber\\
&-&m_{q_2}\frac{A^{-1}_{12}}{2}
{\rm tr} \Bigl[\Gamma^{i}_{2}\Gamma^{f}_{2}\bigl(\nv\alpha_{3}
+\nvpr\alpha_{4}\bigr)\Bigr]
\nonumber\\
&+&\frac{A^{-1}_{12}\Bigl(A^{-1}_{11}+A^{-1}_{12}\Bigr)}{4}
{\rm tr}\Bigl[\Gamma^{i}_{2}\bigl(\nv\alpha_{3}+\nvpr\alpha_{4}\bigr)
\nonumber\\
&\times&\Gamma^{f}_{2}\bigl(\nv\alpha_{3}
+\nvpr\alpha_{4}\bigr)\Bigr] 
\nonumber\\
&-&\frac{A^{-1}_{12}+A^{-1}_{22}}{4}
{\rm tr}\Bigl[\Gamma^{i}_{2}\gamma^{\alpha}\Gamma^{f}_{2}\gamma_{\alpha}
\Bigr]
\nonumber\\
&\times&
\left.
\left(\alpha_{3}\frac{dz}{d\alpha_{3}}+\alpha_{4}\frac{dz}{d\alpha_{4}}
\right)\right\}
\nonumber
\end{eqnarray}
where 

\begin{eqnarray}
z&=&m^2_{q_1}\alpha_1+m^2_{q_2}\alpha_2
-\bar\Lambda_{q_1q_2}\bigl(\alpha_3+\alpha_4\bigr)
\\
&+&\frac{2+\alpha_1+\alpha_2}{4\left|A\right|}
\Bigl[\alpha^2_3+\alpha^2_4+2\alpha_3\alpha_4\omega\Bigr]
\nonumber
\end{eqnarray}

\begin{eqnarray}
A=\left(\barr{ll}
2+\alpha_2  & 1+\alpha_2\\[3mm]
1+\alpha_2  & 2+\alpha_1+\alpha_2\earr\right)
\end{eqnarray}
Here $|A|={\rm det}\{ A \}$.

In the heavy quark limit the matrix elements describing semileptonic 
$b\to c$ transitions 
can be expressed through the three universal Isgur-Wise 
functions $\zeta(\omega)$, $\xi_1(\omega)$ and $\xi_2(\omega)$ of the  
dimensionless variable $\omega=v\cdot v^\prime$ where $v$ and $v^\prime$ are 
the four-velocities of initial and final baryons, respectively.  One finds 

\underline{$\Lambda_{\,b} \longrightarrow \Lambda_{\,c}$ transition}

\beq
\left<\Lambda_{\,c}\bigl(v^{\,\prime}\bigr)\!\left|\,
\overline{b}\,\Gamma \,c\,\right|\!\Lambda_{\,b}\bigl(v\bigr)\right>=
\zeta\bigl(\omega\bigr)\overline{u}\bigl(v^{\,\prime}\bigr)\Gamma
u\bigl(v\bigr),
\eeq

\underline{$\Omega_b \rightarrow \Omega_c(\Omega_c^*)$ transition}
\begin{eqnarray}
&&\left<\Omega_{c}\bigl(v^{\,\prime}\bigr)\,
\mbox{or}\,\Omega_{c}^{\,*}\bigl(v^{\,\prime}\bigr)\!\left|\,
\overline{b}\, \Gamma\, c\,\right|\!\Lambda_{\,b}\bigl(v\bigr)\right>=
\\
&&{\overline{B}}^{\mu}_{\,c}\bigl(v^{\,\prime}\bigr)
\Gamma B^{\,\nu}_{\,b}\bigl(v\bigr)
\bigl[-\xi_1\bigl(\omega\bigr)g_{\mu\nu}
+\xi_2\bigl(\omega\bigr)v_\mu v^{\,\prime}_\nu\bigr]
\nonumber
\end{eqnarray}
where the spinor tensor $\ds{B^{\,\nu}_b(v)}$ obeys the Rarita-Schwinger
constraints  $\ds{v_\nu B^{\,\nu}_b(v)=0}$ and $\gamma_\nu B^{\,\nu}_b(v)=0$.
The spin-wave functions are written as following:
\beq
\barr{llll}
B^{\,\mu}_Q(v)&=&\ds{\frac{\gamma^{\,\mu}+v^{\,\mu}}
{\sqrt{3}}}\gamma^{\,5} u_{\,\Omega_Q}(v)\hfill&
\mbox{for $\Omega_Q$ states}\\[3mm]
B^{\,\mu}_Q(v)&=&u^{\,\mu}_{\,\Omega^{\,*}_Q}(v)\hfill&
\mbox{for $\Omega^{\,*}_Q$  states}
\earr
\eeq
where  the $u_{\,\Omega_Q}\bl(v\br)$ and
$u^{\,\mu}_{\,\Omega^{\,*}_Q}\bl(v\br)$
are the spin-$\ds{1\!\left/2\right.}$
spinor and the Rarita-Schwinger spinor, respectively.

\subsection{Baryonic Isgur-Wise functions}

A direct evaluation of the baryon Isgur-Wise functions with the currents 
(\ref{cur_mix}) gives the following analytical results: 
\begin{equation}
\label{expres}
\zeta\bl(\omega\br)=\frac{F_0\bl(\omega\br)}{F_0\bl(1\br)}\,,
\hspace{0.2cm}
\xi_1\bl(\omega\br)=\frac{F_1\bl(\omega\br)}{F_1\bl(1\br)}\,,
\hspace{0.2cm}
\xi_2\bl(\omega\br)=\frac{F_2\bl(\omega\br)}{F_1\bl(1\br)}
\end{equation}

\begin{eqnarray}
F_{\rm I}(\omega)=
\ds{\int\limits_0^\infty \!dx\,x\!\int\limits_0^\infty}\!
\ds{dy\,y
\int\limits_0^1 \!d\phi\!\int\limits_0^1 \!d\theta
\,\frac{R_I\bl(\omega\br)}{|A|^2}
{\widetilde{F}}^{2}_{B_{Q}}\bl(6z\br)}
\end{eqnarray}
where

\begin{eqnarray}
R_0(\omega)&=&m_{q_1}m_{q_2}(\alpha^2_P+\alpha^2_A\,\omega)
\nonumber\\
&+&\left(2\alpha^2_P-\alpha^2_A\,\omega\right)
\frac{x}{\left|A\right|}
\nonumber\\
&\times&
\left[\frac{2+y}{\left|A\right|}x
\Bigl(1+2\phi(1-\phi)(\omega-1)\Bigr)
-2\bar\Lambda_{q_1q_2}\right]
\nonumber\\
&+&(\omega+1)[m_{q_1}(1+y\theta)
\nonumber\\
&+&m_{q_2}(1+y(1-\theta))]
\alpha_P\alpha_A\frac{x}{\left|A\right|}\,
\nonumber\\
&+&\frac{x^{2}}{4\,|A|^2}\Bigl[\alpha^2_P+\alpha^{2}_{A}\,\omega
\nonumber\\
&+&2\phi\bigl(1-\phi\bigr)\bigl(\omega-1\bigr)
\bigl(\alpha^2_P-\alpha^2_A\bigr)\Bigr] 
\nonumber\\
&\times&
\Bigl[1+y+y^{2}\theta(1-\theta)\Bigr],
\nonumber\\
& &\nonumber\\
R_1(\omega)&=&m_{q_1}m_{q_2}(\beta^2_V+\beta^2_T\,\omega)
\nonumber\\
&+&\beta^2_V\frac{x}{\left|A\right|}\left[\frac{2+y}{\left|A\right|}x
\Bigl(1+2\phi(1-\phi)(\omega-1)\Bigr)
-2{\bar\Lambda}_{q_1q_2}\right]
\nonumber\\
&+&(\omega+1)[m_{q_1}(1+y\theta)
\nonumber\\
&+&m_{q_2}(1+y(1-\theta))] 
\,\beta_V\beta_T\frac{x}{\left|A\right|}\,
\nonumber\\
&+&\frac{x^2}{4\,|A|^2}\Bigl[\beta^2_V+\beta^2_T\,
\omega
\nonumber\\
&+&2\phi(1-\phi)(\omega-1)(\beta^2_V-\beta^2_T)\Bigr]
\nonumber\\
&\times&
\Bigl[1+y+y^2\theta(1-\theta)\Bigr],\nonumber\\
& &\nonumber\\
R_2(\omega)&=&m_{q_1}m_{q_2}\,\beta^2_T+
[m_{q_1}(1+y\theta)
\nonumber\\
&+&m_{q_2}(1+y(1-\theta))]
\,\beta_V\beta_T\frac{x}{\left|A\right|}\,
\nonumber\\
&+&\frac{x^2}{4\,|A|^2}\Bigl[\beta^2_T
+2\phi(1-\phi)(\beta^2_V-\beta^2_T)\Bigr]
\nonumber\\
&\times&
\Bigl[1+y+y^2\theta(1-\theta)\Bigr]\,.
\nonumber 
\end{eqnarray}
Here
\begin{eqnarray}
& &\left|A\right|=3+2y+y^{\,2}\theta\bl(1-\theta\br)\,
\nonumber\\
& &z=y(m_{q_1}^2 \theta + m_{q_2}^2 (1-\theta))\,
\nonumber\\
&+&
x\left[\frac{2+y}{4\left|A\right|}x
\Bigl(1+2\phi\bl(1-\phi\br)\bl(\omega-1\br)\Bigr)
-\bar\Lambda_{q_1q_2}\right]\,.
\nonumber
\end{eqnarray}

\begin{figure}[t]
\epsfig{file=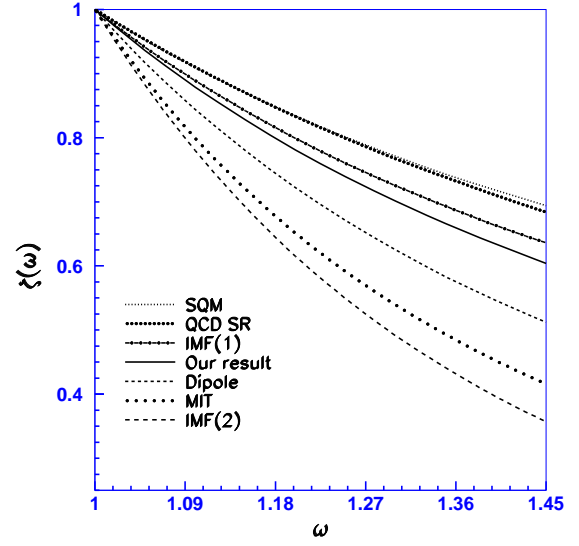,height=8cm}
%   \epsfxsize=12cm
%   \centerline{\epsffile{fig2.eps}}
\caption
{
The  Isgur-Wise function $\zeta(\omega)$ of the decay 
$\Lambda^0_b\to\Lambda^+_c e^-\bar \nu_e$: 
SQM (Simple Quark Model, Ref.~\protect\cite{SQM}); 
QCD SR (QCD Sum Rule, Ref.~\protect\cite{Dai});
IMF(1) (Infinite Momentum Frame Quark Model, 
Ref.~\protect\cite{IMF1}); 
IMF(2) (Infinite Momentum Frame Quark Model, 
Ref.~\protect\cite{IMF2}); 
Our result (for $\bar\Lambda=0.6$ GeV; $\Lambda_{B_Q}=1.8$ GeV, 
$\alpha_A=0$);
Dipole (Dipole form factor, Ref.~\protect\cite{IMF1});
MIT (MIT Bag Model, Ref.~\protect\cite{Sadzik_Zalew}).
}
\label{fig2}
\end{figure}

\begin{figure}[t]
\epsfig{file=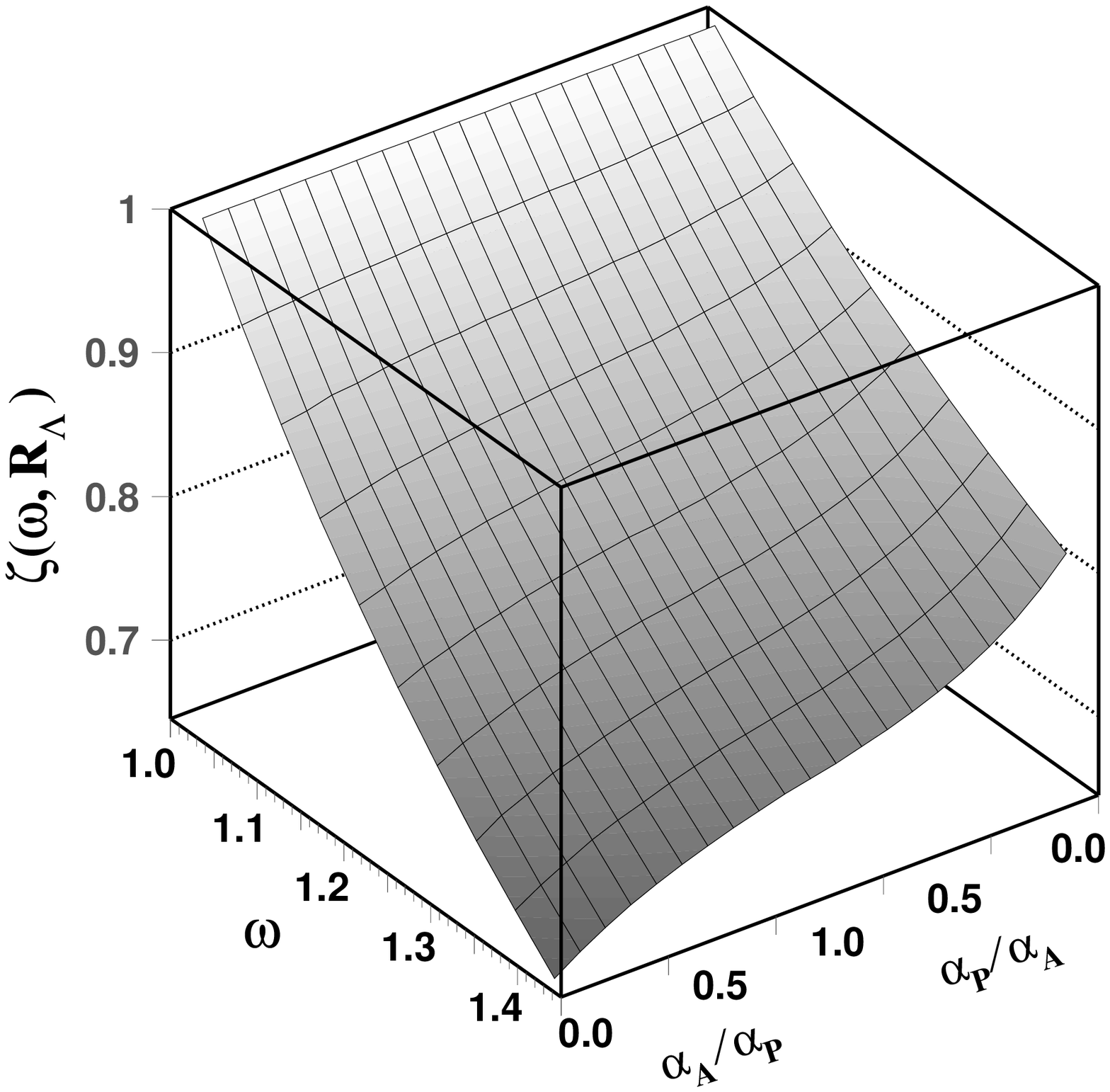,height=8cm}
%   \epsfxsize=12cm
%   \centerline{\epsffile{fig3.eps}}
\caption{
The sensitivity of the Isgur-Wise function $\zeta(\omega)$ 
($\Lambda_b$-decay) on the choice of the three-quark currents 
at fixed values of $\Lambda_{B_Q}$=1.8 GeV and $\bar\Lambda=0.6$ GeV.
}
\label{fig3}
\end{figure}

\begin{figure}[t]
\epsfig{file=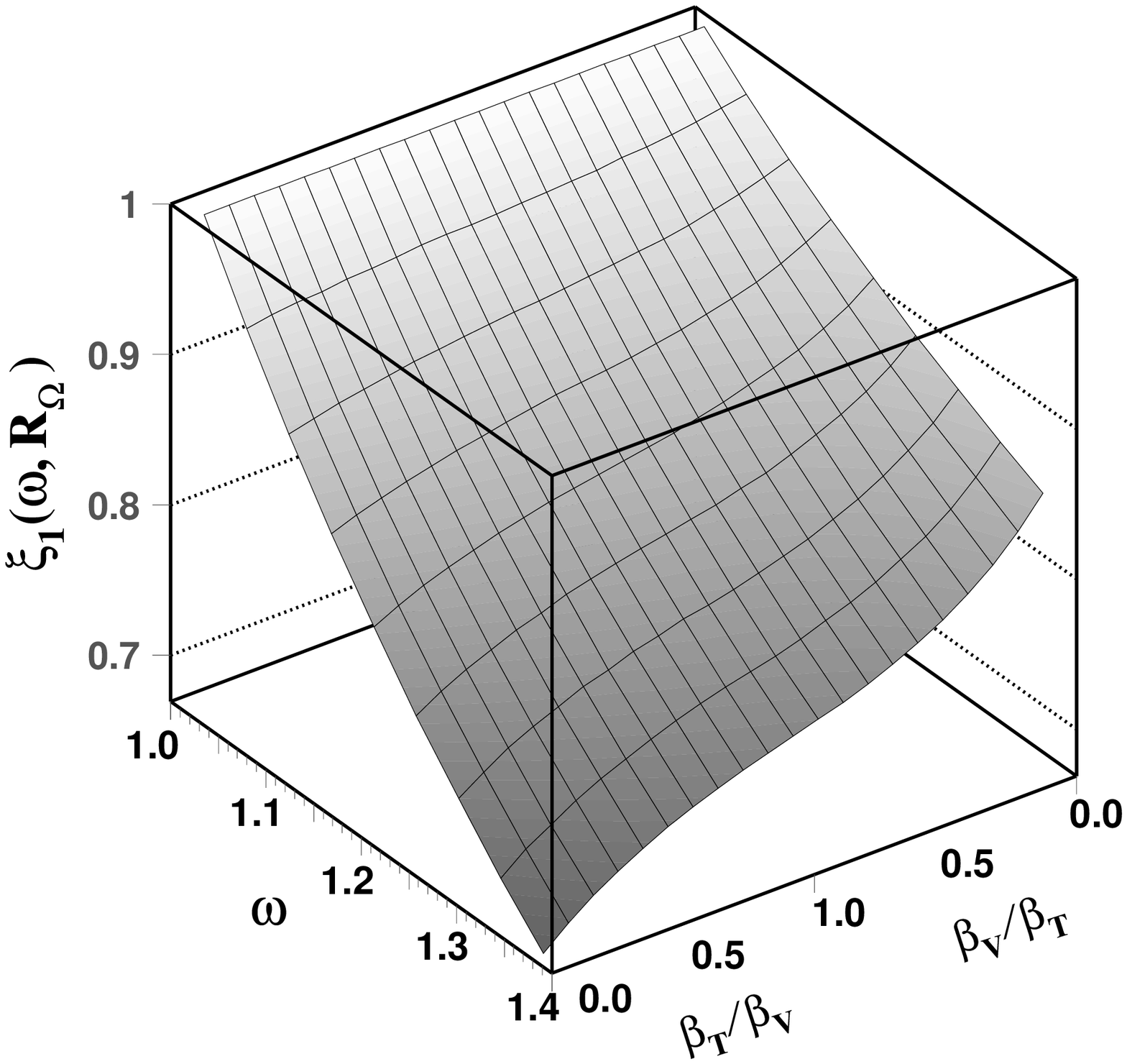,height=8cm}
%   \epsfxsize=12cm
%   \centerline{\epsffile{fig4.eps}}
\caption{
The sensitivity of the  Isgur-Wise  function $\xi_1$ ($\Sigma_{b}$-decay)
on the choice of three-quark currents 
at fixed values of  
$\Lambda_{B_Q}=1.8$ GeV and $\bar\Lambda=0.6$ GeV. 
}
\label{fig4}
\end{figure}

\begin{figure}[t]
\epsfig{file=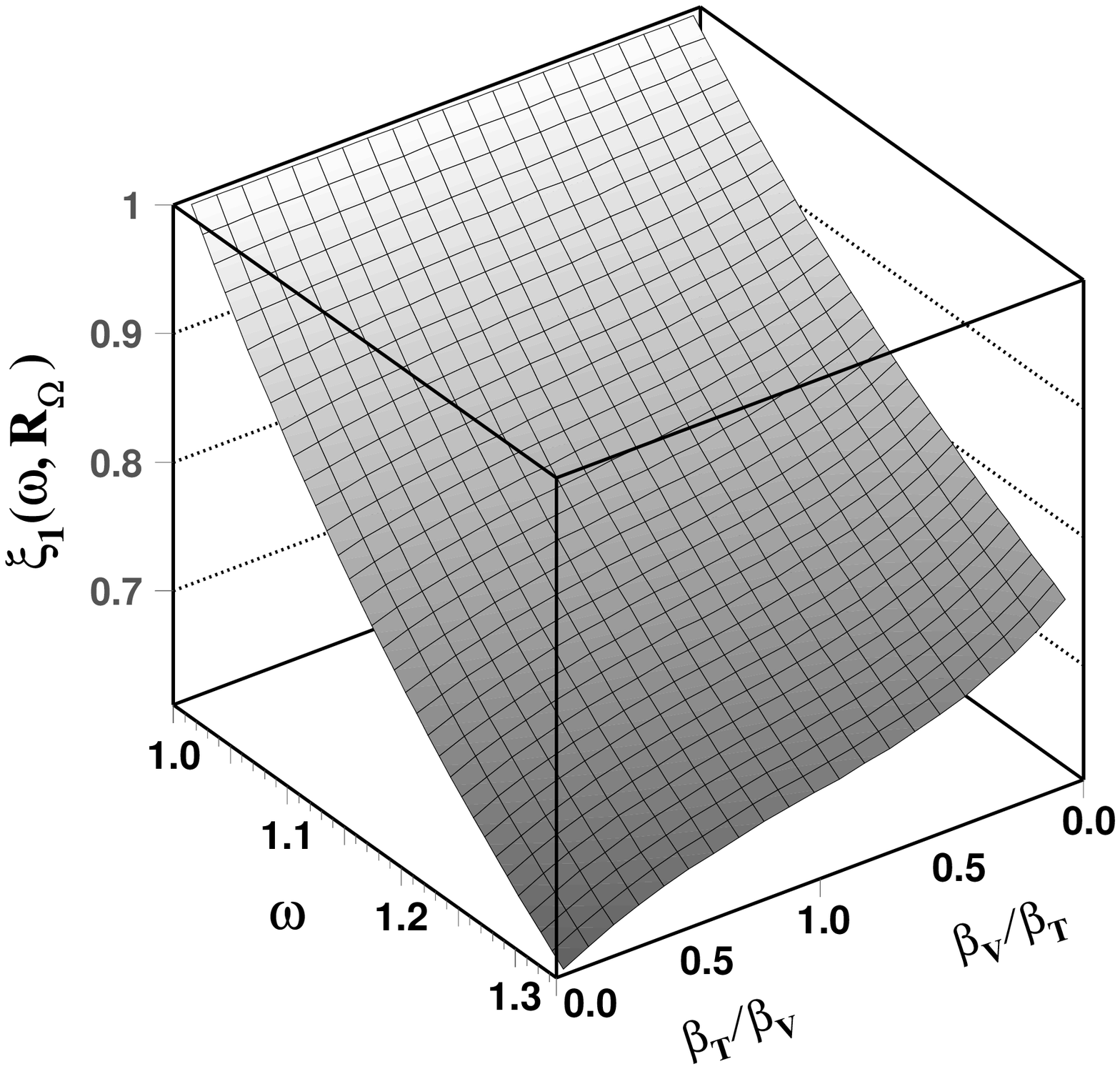,height=8cm}
%   \epsfxsize=12cm
%   \centerline{\epsffile{fig5.eps}}
\caption{
The sensitivity of the Isgur-Wise function $\xi_1$  ($\Omega_{b}$-decay)
on the choice of three-quark currents  at fixed values of  
$\Lambda_{B_Q}=1.8$ GeV and $\bar\Lambda_{ss}=0.9$ GeV. 
}
\label{fig5}
\end{figure}

In Fig.~\ref{fig2} we depict our model predictions for 
the $\omega$-dependence of the form factor $\zeta(\omega)$. 
We compare our prediction for $\zeta(\omega)$  with  results from the simple 
quark model \cite{SQM}, QCD sum rules \cite{Dai}, the infinite
momentum frame quark model \cite{IMF1,IMF2}, the dipole model \cite{IMF1} 
and the MIT bag model \cite{Sadzik_Zalew}.
In Fig.~(\ref{fig3}-\ref{fig5}) we exhibit the  sensitivity of 
the Isgur-Wise functions $\zeta(\omega)$ and $\xi_1(\omega)$ 
to the choice of the three-quark currents with the quantum numbers of 
$\Lambda$-type and $\Omega$-type baryons. We have kept  the values of the  
other model parameters  fixed in this comparison 
($\bar\Lambda$=0.6 GeV, $\bar\Lambda_{\{ss\}}$=0.9 GeV,
and $\Lambda_{B_Q}$=1.8 GeV). 
The Isgur-Wise function $\zeta(\omega)$ for the $\Lambda_b\to\Lambda_c$ 
transition calculated with the pseudoscalar current can be seen to lie
below the one calculated with the axial current. This will result in
different rate predictions.
Similarly, in the case of the $\Omega$-type baryons 
(see, Fig.~\ref{fig4} and Fig.~\ref{fig5}), 
the vector current  Isgur-Wise function lies below  
the tensor current Isgur-Wise function. 

The radii of the form factors $\ds{\zeta}$ and $\ds{\xi_1}$ (or the
slope parameters) are defined by 
\beq\label{radius}
F(\omega)=1-\rho^{\,2}_F(\omega-1)+\ldots
\eeq
where $F=\zeta$ or $\xi_1$. Varying the parameters $R_\Lambda$ and $R_\Omega$ 
in the range $[0,\infty)$  and keeping the values of $\Lambda_{B_Q}$ and 
$\bar\Lambda_{q_1q_2}$ fixed,
the slopes of the  $\Lambda_b$ and $\Sigma_b$ baryon Isgur-Wise 
functions are given by 
$\rho^2_\zeta=1.05 \pm 0.30$ and 
$\rho^2_{\xi_1}=1.07 \pm 0.30$. 
In particular, one has 

\begin{eqnarray}\label{r2}
\begin{array}{ll}
\rho^2_\zeta=1.35   & \hfill \mbox{ for \, $\alpha_A/\alpha_P=0$ }\,, \\[1mm] 
\rho^2_\zeta=1.05   & \hfill \mbox{ for \, $\alpha_P/\alpha_A=1$ }\,, \\[1mm]
\rho^2_\zeta=0.75   & \hfill \mbox{ for \, $\alpha_P/\alpha_A=0$ }\,, \\[1mm]
\rho_{\xi_1}=1.37   & \hfill \mbox{ for \, $\beta_T/\beta_V=0$   }\,, \\[1mm] 
\rho^2_{\xi_1}=1.06 & \hfill \mbox{ for \, $\beta_T/\beta_V=1$   }\,, \\[1mm]
\rho^2_{\xi_1}=0.75 & \hfill \mbox{ for \, $\beta_V/\beta_T=0$   }\,. 
\end{array}
\end{eqnarray}

Finally we cite  the values of the charge radius of the 
$\Lambda_b\to\Lambda_c$ Isgur-Wise function of other theoretical 
model calculations. They vary in a rather large range: 

$$
\rho^{2}_{\zeta}:
$$

\begin{eqnarray}
\begin{array}{ccccccc}
    $\cite{IMF1}$ & $ \cite{Skyrme} $ & $ \cite{Dai} $ & $\cite{SQM}$ &
     $\cite{Sadzik_Zalew}$ & $\cite{Grozin}$ & $\cite{Bowler}$ \\
\hline
% \mbox{$\rho^{2}_{\zeta}$ }
    $\ 1.44 \ $ & $ \ 1.3\   $  &  $ \ 0.65 \ $ & 
      $\ 1.01 \ $ & $ \ 2.35 \ $  &  $ \ 1.15 \ $ & 
      $\  $1.2^{+0.8}_{-1.1}$\  $ 
\\
\end{array}
\nonumber
\end{eqnarray}

\subsection{Rates, distributions and asymmetry parameters}

In this section we present our numerical results on rates, distributions and 
asymmetry parameters for the $b\to c$ flavour changing heavy baryon decays. 
The standard expressions for observables of semileptonic decays of bottom 
baryons (decay rates, differential distributions, leptonic spectra, and 
asymmetry parameters) have simple forms when expressed in terms of helicity 
amplitudes. The set of HQL helicity amplitudes describing transitions of 
bottom baryons into charm baryons can be found in Ref. \cite{ILKK}. 
In Table~\ref{tab1} we present our numerical results for the total and 
partial rates of $\Lambda_b\to\Lambda_c e \nu_e$ transitions using three 
particular choices of the $R_\Lambda$ parameter: $\alpha_A/\alpha_P=0$, 
$\alpha_P/\alpha_A=0$ and $\alpha_P/\alpha_A=1$. One can see that the  
 pseudoscalar current consistently gives  smaller rate values. 
However, the numbers show that the difference  between the three choices 
is not very significant. 
Our results are in good agreement with the experimental 
upper limit for the rate $\Gamma(\Lambda_b\to\Lambda_c e \nu_e)$ 
given by $(6.67\pm 2.73) \times 10^{-10}$ s $^{-1}$. 
For comparison we present the results of some other theoretical approaches. 
In Table~\ref{tab2} we give our predictions for the  other 
semileptonic decay rates of bottom baryons.

\begin{figure}[t]
\epsfig{file=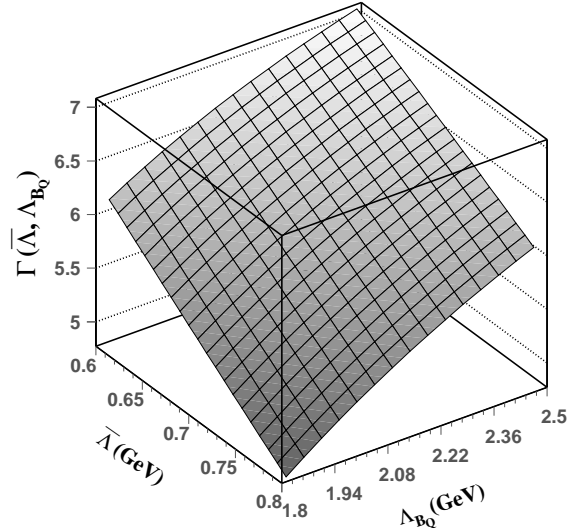,height=8cm}
%   \epsfxsize=12cm
%   \centerline{\epsffile{fig6.eps}}
\caption{
Sensitivity of the total rate of $\Lambda_b\to\Lambda_c+e+\nu_e$ 
transition on the choice of the model parameters $\bar\Lambda$ and 
$\Lambda_{B_Q}$ keeping the  parameter $R_\Lambda$ fixed at $R_\Lambda=1$.   
}
\label{fig6}
\end{figure}

\begin{figure}[t]
\epsfig{file=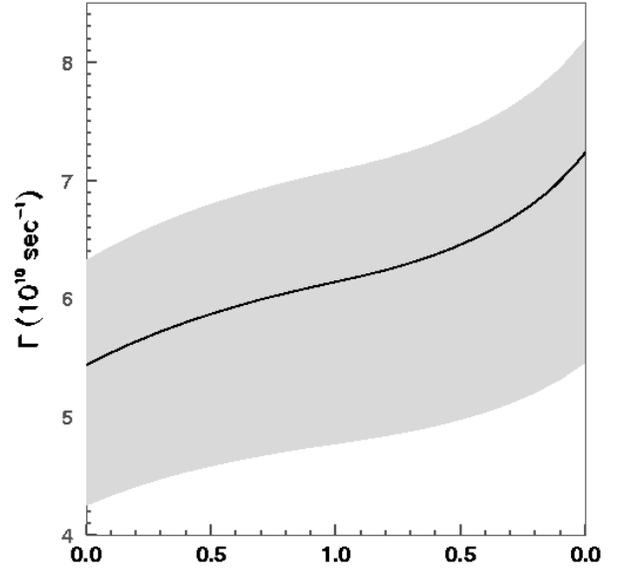,height=8cm}
%   \epsfxsize=12cm
%   \centerline{\epsffile{fig7.eps}}
\caption{
The sensitivity of the $\Lambda_b\to\Lambda_c + e \nu_e$ decay rate 
on the choice of three-quark currents parametrized by the ratio
$R_\Lambda=\alpha_A/\alpha_P$. 
The shaded region corresponds to the range of the total rate
with the model parameters $\bar\Lambda$ and $\Lambda_{B_{Q}}$ being 
varied in the region 
$\ds{0.6\mbox{ GeV}< \bar\Lambda <0.8\mbox{ GeV}}$ and 
$\ds{1.8\mbox{ GeV}<\Lambda_{B_{Q}}<2.5\mbox{ GeV}}$.
The solid heavy curve corresponds to the set 
$\bar\Lambda=0.6$ GeV and $\Lambda_{B_{Q}}=1.8$ GeV. 
}
\label{fig7}
\end{figure}

\begin{figure}[t]
\epsfig{file=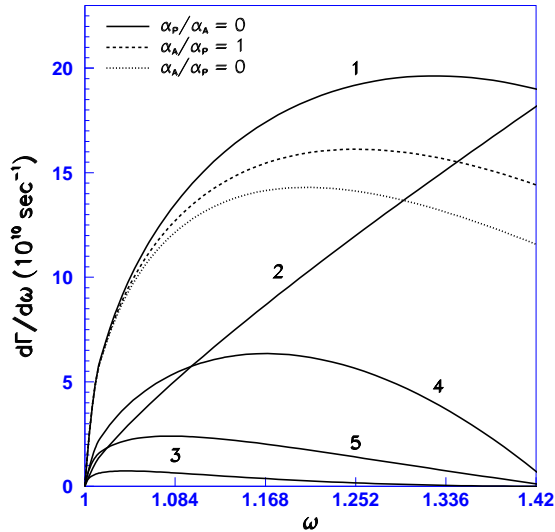,height=8cm}
%   \epsfxsize=12cm
%   \centerline{\epsffile{fig8.eps}}
\caption{
Differential distributions in semileptonic  
$\Lambda_b\to\Lambda_c e \bar\nu_e$ decays for  
$\bar\Lambda=0.6$ GeV; $\Lambda_{B_{Q}}=1.8$ GeV and $V_{bc}=0.04$.
$1$:~$d\Gamma/d\omega.$
$2$:~$d\Gamma_{L_{-}}/d\omega.$
$3$:~$d\Gamma_{L_{+}}/d\omega.$
$4$:~$d\Gamma_{T_{-}}/d\omega.$
$5$:~$d\Gamma_{T_{+}}/d\omega.$
The solid, dashed and dotted line correspond to a set of the parameters 
$\alpha_{A}/\alpha_{P}=0$, $\alpha_{A}/\alpha_{P}=1$ and 
$\alpha_{P}/\alpha_{A}=$, respectively.
}
\label{fig8}
\end{figure}

\begin{figure}[t]
\epsfig{file=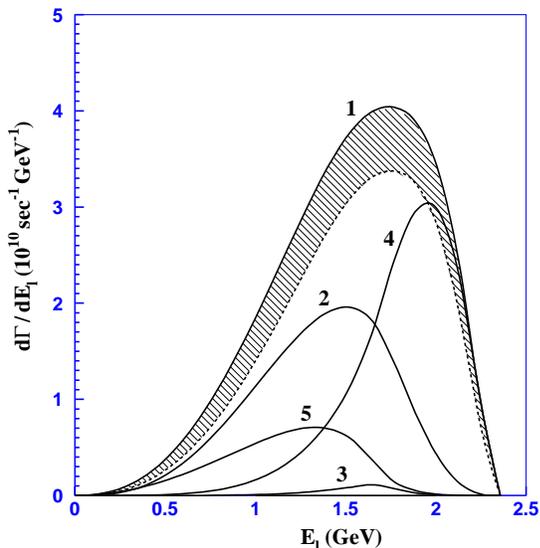,height=8cm}
\caption{\label{fig9}
Leptonic spectrum in the semileptonic decay 
$\Lambda_b\to\Lambda_c e \bar\nu_e$  for 
$\bar\Lambda=0.6$ GeV; $\Lambda_{B_{Q}}=1.8$ GeV and $V_{bc}=0.04$. 
The curve $(1)$ corresponds to $d\Gamma/dE_{\ell}$, 
$(2)$ corresponds to $d\Gamma_{L_{-}}/dE_{\ell}$, 
$(3)$ corresponds to $d\Gamma_{L_{+}}/dE_{\ell}$, 
$(4)$ corresponds to $d\Gamma_{T_{-}}/dE_{\ell}$ and 
$(5)$ corresponds to $d\Gamma_{T_{+}}/dE_{\ell}$. 
The shaded region shows the range of the total energy spectrum 
with the  parameter $\ds{R_{\Lambda}}$ being  varied in the interval 
$\ds{0<R_{\Lambda}<\infty}$. The upper curve corresponds to
$\alpha_{P}=0$ and the lower curve corresponds to
$\alpha_{A}=0$.
}
\end{figure}

In Fig.~\ref{fig6} we depict the dependence of 
the $\Lambda_b\to\Lambda_c e \nu_e$ rate on the parameters $\bar\Lambda$ 
and $\Lambda_{B_Q}$ where the latter parameters are varied in the regions 
0.6 GeV $< \bar\Lambda <$ 0.8 GeV and 1.8 GeV $< \Lambda_{B_Q} <$ 2.5 GeV and 
the current mixing parameter $R_\Lambda$ equals to $R_\Lambda=1$.  
For the rate we find 
$\Gamma(\Lambda_b\to\Lambda_c e \nu_e)=(5.9.\pm 1.1) \times 
10^{10}$ s$^{-1}$ where the theoretical error results from the variation of
the parameters $\bar \Lambda$ and  $\Lambda_{B_Q}$ in the indicated range.
In Fig.~\ref{fig7} we depict the dependence of 
the  rate $\Gamma(\Lambda_b\to\Lambda_c e \nu_e)$ on the current 
mixing parameter $R_\Lambda=\alpha_A/\alpha_P$ 
with the model parameters $\bar\Lambda$ and $\Lambda_{B_{Q}}$ being 
varied in the region 
$\ds{0.6\mbox{ GeV}< \bar\Lambda <0.8\mbox{ GeV}}$ and 
$\ds{1.8\mbox{ GeV}<\Lambda_{B_{Q}}<2.5\mbox{ GeV}}$.
The solid curve corresponds to the set 
$\bar\Lambda=0.6$ GeV and $\Lambda_{B_{Q}}=1.8$ GeV. It it seen that the 
rate  $\Gamma(\Lambda_b\to\Lambda_c e \nu_e)$ changes in the interval  
$(6.2\pm 2) \times 10^{-10}$ s~$^{-1}$. 
Note the remarkable agreement of our predictions with the available upper 
experimental limit for the rate
$\Gamma(\Lambda_b\to\Lambda_c e \nu_e)= 
(6.67\pm 2.73) \times 10^{-10}$ s $^{-1}$. In Fig.~\ref{fig8}  
and Fig.~\ref{fig9} 
we present our results for the  differential 
decay distribution and the lepton spectra in the   
semileptonic decay $\Lambda_b\to\Lambda_c e \bar\nu_e$.  
Finally, in Table~\ref{tab3} 
we give our predictions for 
the asymmetry parameters (for their definitions, see \cite{H})
which can be measured in the two cascade decays of the $\Lambda_b$ baryon. 
For comparison we also present the results of other theoretical approaches.

\begin{table}[t]
\caption{
Total and partial rates of the semileptonic decay
$\Lambda_b^0\to\Lambda_c^+e^-\bar\nu_e$ 
(in $10^{10}~{\mbox{sec}}^{-1}$) for
$\left|V_{bc}\right|=0.04$, $\Lambda_{B_{Q}}=1.8$~GeV, $\bar\Lambda=0.6$~GeV, 
$\bar\Lambda_s=0.75$~GeV, $\bar\Lambda_{ss}=0.9$~GeV.} 
\label{tab1}
\begin{center}
\begin{tabular}{ccccccc}
\multicolumn{2}{c}{Approach} 
& $\Gamma$       
& $\Gamma_{T_+}$ 
& $\Gamma_{T_-}$ 
& $\Gamma_{L_+}$ 
& $\Gamma_{L_-}$ \\
\hline\hline
%%%%%%%%%%%%%%%%%%%%%%%%%%%%%%%%%%%%%%%%%%%%%%%%%%%%%%%%%%%%%
& $\alpha_A/\alpha_P=0$ & 5.43 & 0.52 & 1.53 & 0.11 & 3.27 \\
Our approach 
& $\alpha_A/\alpha_P=1$ & 6.15 & 0.57 & 1.69 & 0.12 & 3.77 \\
& $\alpha_P/\alpha_A=0$ & 7.23 & 0.63 & 1.93 & 0.13 & 4.54 \\
\hline
%%%%%%%%%%%%%%%%%%%%%%%%%%%%%%%%%%%%%%%%%%%%%%%%%%%%%%%%%%%%%
\multicolumn{2}{c}{IMF \cite{IMF1}} 
& 4.28 & 0.41 & 1.20 & 0.09 & 2.58 \\
\hline
\multicolumn{2}{c}{IMF \cite{IMF2}} 
& 4.89 & 0.44 & 1.53 & 0.10 & 2.82 \\
\hline
\multicolumn{2}{c}{Dipole \cite{IMF1}} 
& 5.42 & 0.55 & 1.58 & 0.12 & 3.17 \\
\hline
\multicolumn{2}{c}{QCD Sum Rule \cite{Lee}} 
& 6.16 & 0.43 & 1.86 & 0.10 & 3.77 \\
\hline
\multicolumn{2}{c}{Large $N_c$ \cite{Lee}} 
& 5.51 & 0.34 & 1.45 & 0.09 & 2.63 \\
\hline
\multicolumn{2}{c}{Experiment \cite{PDG}} 
& $<$ 6.67$\pm$ 2.73 &  &  &  &  \\
%%%%%%%%%%%%%%%%%%%%%%%%%%%%%%%%%%%%%%%%%%%%%%%%%%%%%%%%%%%%%
\end{tabular}
\end{center}
\end{table}

\begin{table}[t]
\caption{
Total and partial semileptonic rates of bottom baryons 
(in $10^{10}~{\mbox{sec}}^{-1}$) 
for $\left|V_{bc}\right|=0.04,$ $\Lambda_{B_{Q}}=1.8$~GeV, 
$\bar\Lambda=0.6$~GeV. $T$ and $L$ stand for the transverse
and longitudinal components of the transition and $(\pm)$ denote
the helicity of the daughter baryon.}

\label{tab2}
\begin{center}
\begin{tabular}{ccccccc}\label{tabrateall}
Decay mode & Currents mixing 
& $\Gamma$       
& $\Gamma_{T_+}$ 
& $\Gamma_{T_-}$ 
& $\Gamma_{L_+}$ 
& $\Gamma_{L_-}$ \\
\hline\hline
%%%%%%%%%%%%%%%%%%%%%%%%%%%%%%%%%%%%%%%%%%%%%%%%%%%%%%%%%%%%%
& $\alpha_A/\alpha_P=0$ & 5.98 & 0.59 & 1.64 & 0.13 & 3.62 \\
$\Xi^0_b\to \Xi^+_c e^-\bar{\nu}_e$ 
& $\alpha_A/\alpha_P=1$ & 6.67 & 0.63 & 1.80 & 0.13 & 4.11 \\
& $\alpha_P/\alpha_A=0$ & 7.72 & 0.70 & 2.02 & 0.14 & 4.86 \\
\hline
%%%%%%%%%%%%%%%%%%%%%%%%%%%%%%%%%%%%%%%%%%%%%%%%%%%%%%%%%%%%%
& $\beta_A/\beta_P=0$ & 2.10 & 0.08 & 0.24 & 1.39 & 0.39 \\
$\Sigma^+_b\to \Sigma^{++}_c e^-\bar{\nu}_e$ 
& $\beta_A/\beta_P=1$ & 2.23 & 0.07 & 0.21 & 1.65 & 0.30 \\
& $\beta_P/\beta_A=0$ & 2.51 & 0.07 & 0.18 & 2.03 & 0.23 \\
\hline
%%%%%%%%%%%%%%%%%%%%%%%%%%%%%%%%%%%%%%%%%%%%%%%%%%%%%%%%%%%%%
& $\beta_A/\beta_P=0$ & 1.60 & 0.07 & 0.19 & 1.07 & 0.27 \\
$\Omega^-_b\to \Omega^0_c e^-\bar{\nu}_e$ 
& $\beta_A/\beta_P=1$ & 1.68 & 0.07 & 0.17 & 1.23 & 0.21 \\
& $\beta_P/\beta_A=0$ & 1.86 & 0.06 & 0.15 & 1.48 & 0.17 \\
\hline
%%%%%%%%%%%%%%%%%%%%%%%%%%%%%%%%%%%%%%%%%%%%%%%%%%%%%%%%%%%%%
& $\beta_A/\beta_P=0$ & 3.69 & 0.50 & 1.27 & 0.86 & 1.06 \\
$\Sigma^+_b\to \Sigma^{*++}_c e^-\bar{\nu}_e$ 
& $\beta_A/\beta_P=1$ & 3.72 & 0.52 & 1.31 & 0.94 & 0.95 \\
& $\beta_P/\beta_A=0$ & 3.84 & 0.54 & 1.39 & 1.05 & 0.86 \\
\hline
%%%%%%%%%%%%%%%%%%%%%%%%%%%%%%%%%%%%%%%%%%%%%%%%%%%%%%%%%%%%%
& $\beta_A/\beta_P=0$ & 4.29 & 0.56 & 1.45 & 1.02 & 1.26 \\
$\Omega^-_b\to \Omega^{*0}_c e^-\bar{\nu}_e$ 
& $\beta_A/\beta_P=1$ & 4.33 & 0.58 & 1.50 & 1.12 & 1.13 \\
& $\beta_P/\beta_A=0$ & 4.43 & 0.60 & 1.58 & 1.24 & 1.01 \\
%%%%%%%%%%%%%%%%%%%%%%%%%%%%%%%%%%%%%%%%%%%%%%%%%%%%%%%%%%%%%
\end{tabular}
\end{center}
\end{table}

\begin{table}[t]
\caption{
Asymmetry parameters of semileptonic $\Lambda_{b}$ baryon decay
for $\Lambda_{B_{Q}}=1.8$~GeV, $\bar\Lambda=0.6$~GeV.}
\label{tab3}
\begin{center}
\begin{tabular}{cccccccc}
\multicolumn{2}{c}{Approach} 
& $\alpha$       
& $\alpha^{\prime}$ 
& $\alpha^{\prime\prime}$  
& $\gamma$ 
& $\alpha_P$ 
& $\gamma_P$ \\
\hline\hline
%%%%%%%%%%%%%%%%%%%%%%%%%%%%%%%%%%%%%%%%%%%%%%%%%%%%%%%%%%%%%
& $\alpha_A/\alpha_P=0$ & -0.77 & -0.11 & -0.53 & 0.55 & 0.40 & -0.16 \\
Our approach 
& $\alpha_A/\alpha_P=1$ & -0.78 & -0.11 & -0.55 & 0.54 & 0.41 & -0.16 \\
& $\alpha_P/\alpha_A=0$ & -0.79 & -0.11 & -0.57 & 0.52 & 0.43 & -0.15 \\
\hline
%%%%%%%%%%%%%%%%%%%%%%%%%%%%%%%%%%%%%%%%%%%%%%%%%%%%%%%%%%%%%
\multicolumn{2}{c}{IMF \cite{IMF1}} 
& -0.76 & -0.11 & -0.53 & 0.55 & 0.39 & -0.16 \\
\hline
\multicolumn{2}{c}{Dipole \cite{IMF1}} 
& -0.75 & -0.12 & -0.51 & 0.57 & 0.37 & -0.17 \\
\hline
\multicolumn{2}{c}{QCD Sum Rule \cite{Lee}} 
& -0.83 & -0.14 & -0.57 & 0.48 & 0.38 & -0.17 \\
\hline
\multicolumn{2}{c}{Large $N_c$ \cite{Lee}} 
& -0.81 & -0.15 & -0.53 & 0.50 & 0.34 & -0.19 \\
%%%%%%%%%%%%%%%%%%%%%%%%%%%%%%%%%%%%%%%%%%%%%%%%%%%%%%%%%%%%%
\end{tabular}
\end{center}
\end{table}

\section{Conclusion}

We have employed the relativistic three-quark model in order to test 
the sensitivity  of bottom baryon decay  observables on the choice of the 
three-quark baryon currents. 
We  have found that the semileptonic decay rates are clearly affected
by the choice of currents, whereas the asymmetry parameters show only a very
weak dependence on the choice of currents. We envisage that more precise
data to be expected in the near future would allow one to determine
the appropriate mixture of currents within a given model such as
the relativistic three-quark model.

\section*{Acknowledgments}
M.A.I. and V.E.L. appreciate  the hospitality at Mainz University 
where  this work was completed. 
The visit of M.A.I. to Mainz University was supported by the DFG (Germany)
and the visit of V.E.L. was supported by the Graduiertenkolleg
``Eichtheorien'' (Mainz). 
This work was supported
in part by the Heisenberg-Landau Program
and by the BMBF (Germany) under contract 06MZ865. J.G.K. acknowledges
partial support by the BMBF (Germany) under contract 06MZ865.
A.G.R. acknowledges partial support of the Swiss National Science
Foundation, and TMR, BBW-Contract No. 97.0131  and  EC-Contract
No. ERBFMRX-CT980169 (EURODA$\Phi$NE).


\begin{references}
\bibitem{KKP} J.G. K\"orner, M. Kr\"amer, and D. Pirjol,
Prog. in Part. Nucl. Phys. {\bf 33}, 787 (1994).
\bibitem{MN} M. Neubert, Phys. Rep. {\bf 245}, 259 (1994).
\bibitem{IW} N. Isgur and M. Wise, Nucl. Phys. {\bf B348}, 276 (1991).
\bibitem{HKKG} F. Hussain, J.G. K\"{o}rner, M. Kr\"{a}mer, and G. Thompson,
Z. Phys. {\bf C51}, 321 (1991).
\bibitem{MRR} T. Mannel, W. Roberts, and Z. Ryzak,
Nucl. Phys. {\bf B355}, 38 (1991).
\bibitem{EIKL} G.V. Efimov, M.A. Ivanov, and V.E. Lyubovitskij,\\
Z. Phys. C {\bf 52}, 149 (1991);  
G.V. Efimov, M.A. Ivanov, N.B. Kulimanova, and V.E. Lyubovitskij, 
Z. Phys. C {\bf 54}, 349 (1992).
\bibitem{H} F. Hussain {\it et. al.}, Nucl. Phys. {\bf B370}, 259 (1992).
\bibitem{KM} J.G. K\"{o}rner and M. Kr\"{a}mer,
Phys. Lett. {\bf B275}, 495 (1992); Z. Phys. {\bf C55},  659 (1992).
\bibitem{Yan} T.M. Yan {\it et. al.}, Phys. Rev. {\bf D46}, 1148 (1992).
\bibitem{Grozin} A.G. Grozin and O.I. Yakovlev, 
Phys. Lett. {\bf B285},  254 (1992); {\bf B291}, 441 (1992). 
\bibitem{Cheng} H.-Y. Cheng {\it et al.} Phys. Rev. {\bf D47}, 1030 (1993).
\bibitem{HDH} M.-Q. Huang, Y.-B. Dai, and C.-S. Huang,
Phys. Rev. {\bf D52}, 3986 (1995). 
\bibitem{Groote} S. Groote, J.G. K\"orner and O.I. Yakovlev,
Phys. Rev. {\bf D54}, 3447 (1996).
\bibitem{ILKK} M.A. Ivanov, V.E. Lyubovitskij, J.G. K\"{o}rner and \\
P. Kroll, Phys. Rev. {\bf D56},  348 (1997).
\bibitem{IKLR1} M.A. Ivanov, J.G. K\"{o}rner, V.E. Lyubovitskij, and \\
A.G. Rusetsky,  Phys. Rev. {\bf D57}, 5638 (1998);
Mod. Phys. Lett. {\bf A13}, 181 (1998).
\bibitem{IKLR2} M.A. Ivanov, J.G. K\"{o}rner, V.E. Lyubovitskij, and \\
A.G. Rusetsky, Phys. Lett. {\bf B442}, 435 (1998);
M.A. Ivanov, J.G. K\"{o}rner, and V.E. Lyubovitskij,
Phys. Lett. {\bf B448}, 143 (1999). 
\bibitem{IKLR3} M.A. Ivanov, J.G. K\"{o}rner, V.E. Lyubovitskij, and \\
A.G. Rusetsky, Phys. Rev. {\bf D59}, 074016 (1999).
\bibitem{IKLR4} M.A. Ivanov, J.G. K\"{o}rner, V.E. Lyubovitskij, and \\
A.G. Rusetsky, Phys. Rev. {\bf D60}, 094002 (1999).
\bibitem{TOK} S. Tawfiq, P.J. O'Donnell, and J.G. K\"{o}rner,
Phys. Rev. {\bf D58}, 054010 (1998);
S. Tawfiq and P.J. O'Donnell, Phys. Rev. {\bf D60}, 014013 (1999).
\bibitem{DFNR} H.G. Dosch, E. Ferreira, M. Nielsen, and R. Rosenfeld,
Phys. Lett. {\bf B431}, 173 (1998);
R.S.M. de Carvalho, {\it et al.}, Phys. Rev. {\bf D60}, 034009 (1999). 
\bibitem{Dai} Y.-B. Dai, C.-S. Huang, M.-Q. Huang and
              C. Liu, Phys. Lett. {\bf B387}, 379 (1996).
\bibitem{SQM} B. Holdom, M. Sutherland and J. Mureika,
Phys. Rev. {\bf D49}, 2359 (1994).
\bibitem{IMF1} B. K\"onig, J.G. K\"orner, M. Kramer and P. Kroll, 
Phys. Rev. {\bf D56}, 4282 (1997).
\bibitem{IMF2}X.H. Guo and P. Kroll, Z. Phys. {\bf C59}, 567 (1993).
\bibitem{Sadzik_Zalew} M. Sadzikowski and K. Zalewski, 
Z. Phys. {\bf C 59}, 677 (1993).
\bibitem{Gupta} D. Chakraverty, T. De, B. Dutta-Roy and K.S. Gupta,\\ 
Mod. Phys. Lett. {\bf A 12}, 195 (1997). 
\bibitem{Lee} J.-P. Lee, C. Liu and H.S. Song,
Phys. Rev. {\bf D58}, 014013 (1998).
\bibitem{Skyrme} E. Jenkins, A.V. Manohar and  M.B. Wise, 
Nucl. Phys. {\bf B396}, 38 (1993).
\bibitem{Bowler}K.C. Bowler {\it et al.,} UKQCD Collaboration, 
Phys. Rev. {\bf D57}, 6948 (1998).
\bibitem{EIL} G.V. Efimov, M.A. Ivanov, and V.E. Lyubovitskij,\\
Few-Body Syst. {\bf 6}, 17 (1989).
\bibitem{EI} G.V. Efimov and M.A. Ivanov, Int. J. Mod. Phys. {\bf A4},
2031 (1989); 
"The Quark Confinement Model", IOP, 1993.
\bibitem{AIKL} I.V. Anikin, M.A. Ivanov, N.B. Kulimanova, and \\
V.E. Lyubovitskij, 
Z. Phys. C {\bf 65}, 681 (1995).
\bibitem{ILL}M.A. Ivanov, M.P. Locher, and V.E. Lyubovitskij,\\
Few-Body Syst. {\bf 21}, 131 (1996).
\bibitem{SWH} A. Salam, Nuovo Cim. {\bf 25}, 224 (1962);
S. Weinberg, Phys. Rev. {\bf 130}, 776 (1963);
K. Hayashi {\it et al.}, Fort. Phys. {\bf 15}, 625 (1967).
\bibitem{WGL} S. Weinberg, Physica {\bf 96A}, 327 (1979);
J. Gasser and H. Leutwyler, Ann. Phys. (N.Y.) {\bf 158}, 142 (1984);
Nucl. Phys. {\bf B250}, 465 (1985).
\bibitem{BL} T. Becher and H. Leutwyler, Eur. Phys. J. {\bf C9}, 643 (1999). 
\bibitem{Shuryak} E.V. Shuryak, Nucl. Phys. {\bf B198}, 83 (1982). 
\bibitem{PDG} C. Caso {\it et.al. }, Eur. Phys. J.  {\bf C3}, 1 (1998).
\end{references}
\end{document}